\documentstyle [12pt,twoside]{article}

\begin{document}
\small
\par\noindent
to appear in: {\it Annals of the New York Academy of Sciences}
\vskip .3in
\centerline{\Large\bf Collisionless Relaxation in Galactic Dynamics}
\vskip .15in
\centerline{\Large\bf and the Evolution of Long Range Order}
\vskip .2in
\renewcommand{\thefootnote}{\alph{footnote}}
\centerline{HENRY E. KANDRUP\footnote{\small
HEK was supported in part by National Science Foundation Grant No.
PHY92-03333. Some of the numerical calculations were facilitated
by computer time provided by {\it IBM} through the Northeast Regional Data
Center (Florida). This article is based in part on collaborations with
Elaine Mahon, Christos Siopis, and Barbara Eckstein, students at the 
University of Florida who have been supported by NASA through the Florida 
Space Grant Consortium.}}
\vskip .1in
\centerline{\it Department of Astronomy and Department of Physics}
\vskip .02in
\centerline{\it and Institute for Fundamental Theory}
\vskip .02in
\centerline{\it University of Florida, Gainesville, Florida 32611}
\vskip .3in
\begin{abstract}
\noindent\small{
This talk presents a critical assessment of certain aspects of collisionless 
galactic dynamics, focusing on the interpretation and limitations of the 
collisionless Boltzmann equation and the physical mechanisms associated with 
collisionless, or near-collisionless, relaxation. Numerical and theoretical
arguments are presented to motivate the idea that the evolution of a
system far from equilibrium should be interpreted as involving nonlinear 
gravitational Landau damping, a picture which implies a greater overall 
coherence and remembrance of initial conditions than is implicit in the 
conventional paradigm of violent relaxation.}
\end{abstract}
\vskip .2in
\centerline{\bf INTRODUCTION AND MOTIVATION}
\vskip .15in
The principal aim of the work described here is to understand the gravitational
$N$-body problem for a collection of nearly point ``particles'' of
comparable mass, such as, for example, the stars in a galaxy. This $N$-body
problem is not directly applicable for problems involving galaxy formation,
or in understanding the structure and evolution of spiral galaxies, where
dissipative gas dynamics must play an important role. However, it should be
relevant in understanding the structure of elliptical galaxies, gas-poor$^{1}$
objects which are typically assumed to be in or near a 
(meta-)equilibrium.$^{2}$ It should, moreover, be important in understanding 
the evolution of elliptical galaxies when, as must often occur,$^{3}$ they 
are displaced from their near-equilbrium by a collision or other close 
encounter with some companion object.
\par
The conventional wisdom of galactic dynamics$^{2}$ asserts that, over 
sufficiently
short time scales, the structure and evolution of a collection of $N$ 
self-gravitating point masses can be described by the {\it collisionless
Boltzmann equation}, the gravitational analogue of the Vlasov-Poisson system
of plasma physics. Alternatively, it is assumed that, on longer time scales,
one must allow for discreteness effects, 
described by the {\it collisional Boltzmann}, or {\it Fokker-Planck equation},
the latter being simply the gravitational analogue of the Landau equation from
plasma physics.
\par
In this context, it is typically assumed (cf. Ref. [4]) that there are four 
distinct sorts of ``relaxation'' processes, namely (1) ``violent relaxation,'' 
(2) gravitational Landau damping, (3) phase mixing, and (4) ``collisional 
relaxation.'' Violent relaxation$^{2,5}$ refers to the early stages of 
evolution 
in which a system, initially far from any steady state, evolves towards some 
meta-equilibrium, i.e., a  time-independent solution to the collisionless 
Boltzmann equation ({\it CBE}). By contrast, Landau damping and phase mixing 
reflect smaller amplitude effects acting in a system which is much closer to 
a meta-equilibrium. Collisional relaxation is associated with discreteness 
effects which, dating back to Chandrasekhar,$^{6}$ have typically been modeled 
as an incoherent sum of binary encounters, i.e., gravitational Rutherford 
scattering.

The first three of these mechanisms, incorporated in the {\it CBE}, are 
presumed to induce an efficient approach towards some meta-equilibrium on a 
time scale
of order $t_{cr}$, a characteristic crossing time for the system in question.
By contrast, collisional relaxation is presumed to be important only on a 
much longer relaxation time $t_{R}$ which, for systems as large as an entire
galaxy, is orders of magnitude longer than $t_{H}$, the age of the Universe.
In the language of statistical physics, one thus assumes that a mean field
theory, based on the {\it CBE}, accurately describes the structure and 
evolution of a gas-poor elliptical galaxy over time scales shorter than or
comparable to the age of the Universe. 

In this talk, I would like to play devil's advocate and ask to what extent
one can really be confident that this conventional wisdom is completely 
correct. Specifically, this talk will address four distinct questions: (1)
Does one really know that the {\it CBE} is the whole story? (2) In what
sense, and to what extent, does the {\it CBE} imply an evolution towards some
meta-equilibrium? (3) How should one interpret the flow associated with the
{\it CBE} and, related to this, what is the dynamical principle that determines
the form of the meta-equilibrium towards which an $N$-body system is presumed
to evolve? (4) What precisely is the connection between the {\it CBE} and the
$N$-body problem, and, in particular, what do {\it CBE} characteristics 
computed for a smooth initial distribution function have to do with orbits
generated as a solution to the full gravitational $N$-body problem?

The principal substantive conclusion is that, contrary to what is usually 
assumed by galactic dynamicists, phase mixing, Landau damping, and violent
relaxation should be viewed as (not so different) manifestations of a single
phenomenon. As is well known to plasma physicists, linear Landau damping 
{\it is} in fact a type of phase mixing. Moreover, numerical experiments and 
theoretical considerations suggest that the evolution of a system far from 
equilibrium can be interpreted fruitfully as entailing nonlinear Landau
damping, this implying a greater overall coherence and remembrance of initial 
conditions than is implicit in the conventional picture of violent 
relaxation.\footnote{\small
It is ironic that Lynden-Bell's original paper$^{5}$ on violent relaxation 
does not implement clear cut distinctions amongst violent relaxation, Landau 
damping, and phase mixing. Rather, it suggests a more integrated approach 
which views phase mixing-cum-Landau damping
as an important part of violent relaxation, an 
approach more consistent, in some respects, with the ideas advocated in this 
paper than with conventional wisdom.}

\vskip .2in
\centerline{\bf IS THE COLLISIONLESS BOLTZMANN EQUATION}
\vskip .05in
\centerline{\bf THE WHOLE STORY?}
\vskip .15in
The basic assumption underlying the {\it CBE} is that the system in question
can be described probabilistically in terms of a smooth one-particle 
distribution function, $f(x^{a},v_{a},t)$, which can be interpreted as a
phase space number or mass density that yields the probability of finding
a particle near the spatial point $x^{a}$ with velocity $v_{a}$ at time $t$.
(In what follows, it will be assumed that each particle has a mass $m=1$ since,
in the context of a mean field theory, allowing for a distribution of particle 
masses has absolutely no effect.) In the context of the initial value problem,
this implies that the true initial distribution function, given 
as a sum of $N$ phase space delta functions, has been replaced by a smooth
function $f(x^{a},v_{a},0)$.
\par
The {\it CBE} manifests the idea that the distribution function $f$ will free 
stream in the self-consistent potential ${\Phi}$ associated with $f$, so that
$${df\over dt}{\;}{\equiv}{\;}{{\partial}f\over {\partial}t}+
v^{a}{{\partial}f\over {\partial}x^{a}}-
{{\partial}{\Phi}\over {\partial}x^{a}}{{\partial}f\over {\partial}v_{a}}=0,
\eqno(1)$$
where 
$${\nabla}^{2}{\Phi}(x^{a},t)=4{\pi}G\,\int\,d^{3}v\,f(x^{a},v_{a},t). 
\eqno(2)$$
In the context of a many-particle phase space description, this is equivalent
to assuming that the two-particle distribution function $g(1,2)$ factorises
into a product of one-particle $f$'s, so that 
$g(1,2){\;}{\approx}{\;}f(1)f(2)$, i.e., neglecting entirely the effects of
particle-particle correlations. It is assumed that this neglect is justified
on time scales short compared with the natural time scale $t_{R}$ associated
with gravitational Rutherford scattering, a time scale which, for very large 
$N$, is much larger than $t_{cr}$.
\par
Unfortunately, this neglect 
of particle correlations constitutes an uncontrolled approximation. Thus, for
example, unlike the case of a neutral plasma, one cannot derive the next order 
correction to the {\it CBE} in the context of some systematic perturbation
expansion. Physically, the problem is that there is no shielding to vitiate
the long range $1/r^{2}$ force. This is, e.g., manifested by the fact that 
a $1/r$ potential yields an infinite cross section, so that, when evaluating
the effects of binary encounters in the usual way for an infinite and 
homogeneous system, one encounters logarithmic divergences in the limit of 
large impact parameter (the problematic Coulomb logarithm of galactic 
dynamics).
\par
Physically, one might hope to circumvent this difficulty by first
identifying the bulk mean field force acting at any given point in space
and then treating ``fluctuations away from mean field conditions'' as the 
source of deviations from mean field theory. For $N{\;}{\gg}{\;}1$ one might 
expect that these fluctuations
are small, so that their effects do in fact constitute a small perturbation.
This splitting into mean field plus fluctuations can be introduced formally,
e.g., by using time-dependent projection operators,$^{7}$ but is difficult to 
implement concretely because of the apparent absence of a clean separation
of time scales. 
\par
One case where one can try to perform a concrete calculation is for the toy 
model of a system that is rotating with an angular velocity ${\Omega}$ 
carefully chosen to ensure that the mean field equilibrium is homogeneous 
(here the rotation provides a fictitious force that plays the role of the 
uniform oppositely charged background in a plasma). In this case, one finds 
that the screened potential $\exp(-{\kappa}r)/r$ of plasma physics is replaced 
by an oscillatory potential $\exp (i{\kappa}r)/r$, this reflecting the fact 
that an homogeneous system is unstable towards sufficiently long wavelength 
perturbations (the Jeans instability). The conventional response is to 
introduce by hand a cut off at a length scale $R_{sys}$ comparable to the size 
of the system, supposing that the system was generated at some earlier time 
by the fragmentation of a much larger, nearly homogeneous entity that 
experienced the Jeans instability. 
\par
As demonstrated by Thirring and coworkers (cf. Ref. [8]), there {\it does} in 
fact exist a well defined thermodynamic limit in which gravitational mean 
field theory becomes exact. However, this limit is an unusual one. Given that
realistic self-gravitating systems are inhomogeneous, one cannot simply
introduce the ordinary thermodynamic limit, with number $N\to\infty$ and
volume $V\to\infty$ but finite density $n=N/V$. Rather, one must instead allow 
for a limit in which the coupling constant $G$ also scales. The problem, 
however, is that there is no guarantee that this limit is physically realistic 
and that, even assuming that it is physical, there are no rigorous (or even
quasi-rigorous) estimates as to the time scale $t_{R}(N)$ on which the mean
field description would be expected to fail.
\par
Given that theoretical analyses have as yet proven inconclusive, one might 
instead seek recourse to numerical experiments. This, however, is difficult. 
For generic systems not characterised by a high degree of symmetry, the 
{\it CBE} is a partial differential equation in six independent phase space 
variables which is extremely expensive to solve computationally. Moreover,
the time required to solve the $N$-body problem using an honest direct 
summation code scales as $N^{2}$, so that the consideration of very large $N$, 
where the {\it CBE} is expected to be valid, becomes prohibitive. The
approximately validity of the {\it CBE} has been corroborated in the sense 
that $N$-body realizations of meta-equilibria often appear to behave stably
for relatively short time scales, and that, at least for short time scales,
various velocity moments extracted from $N$-body simulations and 
time-dependent solutions to the {\it CBE} yield no gross discrepancies. 
However, computational limitations have prevented truly detailed comparisons 
of $N$-body and {\it CBE} evolutions hitherto.
\par
There is in fact one concrete setting where detailed computations have 
been done, namely the toy model of one-dimensional gravity, i.e., a collection
of infinite plane sheets characterized by a two-body potential$^{9}$
$$V(|x_{i}-x_{j}|)=G|x_{i}-x_{j}|. \eqno(3) $$
For this simple model, the phase space is only two-dimensional and the forces
are trivial computationally, so that it is relatively easy to perform highly
accurate computations. The net result of such simulations is that, at least
for times as long as $10^{2}-10^{3}t_{cr}$, there is a reasonably good overall
agreement between $N$-body simulations and an evolution governed by the 
{\it CBE}.$^{10}$  Interestingly, however, the evolution
seems to be substantially more complex than what most astronomers would
expect for three-dimensional gravity (cf. Refs.~[11-12]). These conclusions 
seem highly suggestive. However, they must both be taken with a fair grain of 
salt since there is no compelling reason to believe that the evolution of 
one-dimensional gravitational systems is qualitatively similar to the 
evolution of ``real'' three-dimensional systems.
\par
In summary, even though a mean field theory based on the {\it CBE} may seem 
well motivated physically, there is as yet no rigorous proof of its validity 
and, in particular, no rigorous estimate as to the time scale on which it 
might be expected to fail. Moreover, as discussed below, there are reasons to 
suspect that even very small corrections could be important by accelerating
the approach towards equilibrium and, especially, by serving to ``fuzz out''
small scale structures predicted by the {\it CBE} to arise as a 
self-gravitating system approaches some meta-equilibrium.
\vskip .2in
\centerline{\bf IN WHAT SENSE DOES THE COLLISIONLESS BOLTZMANN}
\vskip .05in
\centerline{\bf EQUATION IMPLY AN APPROACH TOWARDS EQUILIBRIUM?}
\vskip .15in
In addressing the question of evolution, the most important thing to recognise
about the {\it CBE} is that it, like the true $N$-body problem, is Hamiltonian.
Specifically, the {\it CBE} constitutes a non-canonical Hamiltonian system, 
formulated in an infinite-dimensional phase space, where the fundamental
dynamical variable is the distribution function $f$ itself.$^{13,14}$ 
The proof that the {\it CBE} is Hamiltonian entails the identification of a
Hamiltonian functional ${\cal H}[f]$ and a Lie bracket $[\,.\,,\,.\,]$
defined on functionals ${\cal A}[f]$ and ${\cal B}[f]$, so chosen that the
{\it CBE} takes the form
$${{\partial}f\over {\partial}t}+[{\cal H},f] = 0. \eqno(4) $$
\par
It is straightforward to verify that the antisymmetric operation
$$[{\cal A},{\cal B}]=\int\,d^{3}xd^{3}v\,f\,{\Bigl\{}
{{\delta}{\cal A}\over {\delta}f},{{\delta}{\cal B}\over {\delta}f}{\Bigr\}},
\eqno(5) $$
with ${\delta}/{\delta}f$ a functional derivative and $\{ \,.\,,\,.\,\} $
the ordinary Poisson bracket of particle mechanics, defines a {\it bona fide}
Lie bracket 
acting in the infinite-dimensional phase space of distribution functions.
However, if one identifies ${\cal H}[f]$ as the mean field energy, i.e.,
$${\cal H}[f]={1\over 2}\,\int\,d^{3}xd^{3}v\,v^{2}\,f(x^{a},v_{a},t)-
{G\over 2}\;\int\,d^{3}x\,d^{3}v\;\int\,d^{2}x'\,d^{3}v'\;
{f(x^{a},v_{a},t)f(x'^{a},v'_{a},t)\over |{\bf x}-{\bf x}'|}, \eqno(6) $$
it is easy to see that, with this bracket, eq. (4) reduces to the {\it CBE} 
in the form
$${{\partial}f\over {\partial}t}-\{ E,f \} =0 , \eqno(7) $$
where $E={1\over 2}v^{2}+{\Phi}[f]$ is the particle energy. 
\par
Associated with this Hamiltonian character is the existence of an infinite 
number of conserved quantities, the so-called Casimirs, which generalise the 
notion of conservation of phase (or Liouville's Theorem) for a distribution 
function $f$ evolving in an external potential.$^{15}$ Specificially, one
knows that, for any function ${\chi}(f)$, the numerical value of 
$C[f]=\int\,d^{3}xd^{3}v\,{\chi}(f)$ is independent of time.\footnote{
Significantly, however, even though the {\it CBE}
admits this infinite collection of conserved quantities, it is extremely
unlikely that it is integrable (unlike, e.g., the Korteweg-de Vries
equation).}  One concrete manifestation of these Casimirs is the fact that 
any phase-preserving perturbation of a {\it CBE} equilibrium must be generated 
by a canonical transformation.$^{16}$
Every time-independent equilibrium solution $f_{0}$ 
corresponds to an extremal point of the mean field energy, i.e., 
${\delta}^{(1)}{\cal H}{\;}{\equiv}{\;}0$ for all phase preserving 
perturbations of the equilibrium.
\par
The Hamiltonian character of the evolution is crucial because it precludes 
entirely the possibility of any pointwise approach towards a time-independent 
equilibrium. Only by replacing the true distribution function $f$ by some
coarse-grained ${\overline f}$ can one hope to identify an object which 
actually approaches equilibrium. Any meaningful discussion of ``the
approach towards equilibrium'' must be formulated in the context of a 
coarse-grained distribution function or, alternatively, some set of 
coarse-grained observables constructed from the true distribution function 
by some phase space averaging.
\par
Very little is known mathematically about an evolution governed by the 
{\it CBE}. Recently, however, it has been proven that the {\it CBE} manifests
global existence.$^{17,18}$ Specifically, one knows that, given sufficiently
smooth initial data ($C^{1+}$), the distribution function will never diverge.
In other words, unlike the case of a perfect fluid, the evolution never leads
to caustics or shocks. This suggests in particular that smooth initial 
conditions cannot evolve near-singular cores or other lumpy exotica of the
form discovered recently using the Hubble Space Telescope.$^{19}$

Unfortunately, however, there are no rigorous results about a (suitably
coarse-grained) approach towards equilibrium. Rather, the only results obtained
to date relate to the asymptotic behaviour of time integrals of quantities 
like ${\cal K}[f]$ or ${\cal V}[f]$, the mean field kinetic and potential 
energies. Thus, e.g., in certain cases one can prove$^{20}$ the existence of 
quantities like
$$K{\;}{\equiv}{\;}\lim_{T\to\infty}\,{1\over T}\,\int_{0}^{\infty}\;
{\cal K}[f(t)]dt. \eqno(8) $$
However it has not been possible to prove statements about the behaviour of 
(say) ${\cal K}[{\overline f}]$ at a fixed instant of time, let alone the 
behaviour of the full ${\overline f}$.

One might naively suppose that this simply reflects the fact that any such 
result would be very hard to prove. However this
is not completely true. Specifically, there exist exact, time-dependent
solutions to the {\it CBE}, corresponding to a system that remains bounded
in space, that exhibit no approach towards a coarse-grained 
equilibrium.$^{21,22}$ Rather, these solution correspond to finite amplitude,
undamped oscillations about an otherwise time-independent equilibrium $f_{0}$.

There is of course no reason to expect {\it a priori} that such oscillating
solutions actually exist in nature. However, it appears that they can arise
naturally at least in the context of one-dimensional gravity.$^{10}$ 
Specifically, for both $N$-body simulations and the {\it CBE}, the evolution 
of counter-streaming initial conditions (two equilibria engaged in a head-on
collision) can yield a final $f(t)$ which involves undamped oscillations 
about a time-independent $f_{0}$ which, significantly, contains ``holes,''
i.e., regions in the middle of the occupied phase space region where 
$f_{0}\to 0$.
\vskip .2in
\centerline{\bf HOW SHOULD ONE INTERPRET THE FLOW ASSOCIATED}
\vskip .05in
\centerline{\bf WITH THE COLLISIONLESS BOLTZMANN EQUATION?}
\vskip .15in
\centerline{\bf ``Violent Relaxation''}
\vskip .1in
Most discussions of the evolution of a self-gravitating system initially far 
from a meta-equilibrium have been formulated within the context of the 
paradigm of violent relaxation, a sort of ``self-consistent egg beater'' model
originally proposed by Lynden-Bell thirty years ago.$^{5}$ Different workers
have interpreted this paradigm in rather different ways. Indeed, the 
conventional wisdom would appear to ignore many of the fine points made in
the original paper. However, (almost) all 
versions of violent relaxation seem to incorporate four basic ingredients:
\begin{enumerate}
\item Attention focuses on a coarse-grained distribution function 
${\overline f}$ which can ``fuzz out'' in a way that is impossible for the
true $f$, which is strongly constrained by conservation of phase.
The obvious point here is that if $f$ satisfies the {\it CBE}, in
general the coarse-grained ${\overline f}$ will not.
Lynden-Bell starts from the assumption that the fundamental object of interest
is a smooth one-particle $f$ that satisfies exactly the {\it CBE} and then
implements a coarse-graining of the one-particle phase space to extract an
appropriate ${\overline f}$. However, many later workers
have sought to relate ${\overline f}$ directly to the full $N$-particle
dynamics, as derived, e.g., from numerical simulations. This difference in
perspective is important. Lynden-Bell assumes that the {\it CBE} is exact in
some suitable $N\to\infty$ continuum limit and formulates his entire theory
in that limit. For him, therefore, violent relaxation need say little if 
anything about individual particles, and should be interpreted instead as 
tracking a coarse-grained free streaming of phase space fluid elements.
\item 
\vskip -.15in 
One invokes the effects of a strongly time-dependent bulk potential
${\Phi}[f]$ to help ``shuffle'' the coarse-grained ${\overline f}$.
The initial $f(0)$ yields the mass density ${\rho}(0)$ and, hence, the 
gravitational potential ${\Phi}(0)$ which determines the initial motion of 
each phase point in the system. In general, $f(0)$ will be far from 
equilibrium, so that ${\Phi}$ will induce changes in $f$ on a time scale
${\sim}{\;}t_{cr}$. However, changes in $f$ correspond in general to changes
in ${\Phi}$, so that the potential itself will vary significantly on the 
same time scale $t_{cr}$. This implies in turn that the particle energy
$E={1\over 2}v^{2}+{\Phi}$ associated with any characteristic will be strongly 
time-dependent. In particular,
if the potential exhibits large enough variations on a sufficiently short
time scale, $E$ will be far from an adiabatic invariant, so that one might
expect that the energies of the different characteristics will become ``well 
shuffled'' and/or randomized. 
\item 
\vskip -.15in
One supposes that, as a result of this convulsive process, ${\overline f}$ 
will evolve towards, albeit not necessarily to, the ``most likely'' 
${\overline f}_{0}$ consistent with 
an appropriate set of constraints. Two different types of constraint are 
usually considered, namely (1) bulk holonomic constraints, such as conservation
of energy and particle number, and (2) the residual, coarse-grained effects
of conservation of phase, which may still prove important if the system is
sufficiently dense. Although crucial conceptually, the latter are often 
unimportant as a practical matter (unless, e.g., the system is very dense)
and, for that reason, will be ignored in the following.
\item
\vskip -.15in
One supposes further that the ``most likely'' ${\overline f}_0$ can be derived 
from a combinatoric argument by determining the macrostate corresponding to 
the largest number of microstates. To the extent that the constraints
associated with conservation of phase are unimportant, this leads to the
identification of the most likely ${\overline f}_{0}$ as being that particular 
coarse-grained distribution function that maximizes the Boltzmann entropy
$$S[{\overline f}]=-{\int}\,d^{3}x\;\int\,d^{3}v{\;}
{\overline f}\,{\rm log}\,{\overline f}. \eqno(9)$$
\end{enumerate}
\par\noindent  
It is easily seen that, by combining these four ingredients, one is led to a
predicted evolution towards an isothermal distribution of the form
$${\overline f}_{0}{\;}{\propto}{\;}
\exp {\bigl\{}-{\beta}(v^{2}/2 + {\Phi}[{\overline f}_{0}])\,
{\bigr\}}. \eqno(10) $$
(Allowing for coarse-grained conservation of phase would modify eq. (9) to
yield a degenerate isothermal.)
\par
Lynden-Bell would stress that, albeit involving combinatoric arguments familiar
from information theory, his derivation of eq.~(10), which deals with phase
space fluid elements rather than discrete particles, is not an entropy
argument in the usual sense. One clear indication of this is the fact that
the particle mass $m$ does not enter into ${\overline f}_{0}$. If one 
implements a ``true'' entropy argument for a collection of particles with 
a distribution of masses, one finds instead$^{64}$ a mass-dependent extremal 
entropy configuration of the form
${\overline f}_{0}{\;}{\propto}{\;}
\exp {\bigl\{}-{\beta}(mv^{2}/2 + m{\Phi}[{\overline f}_{0}])\,{\bigr\}}.  $
\par
The ``prediction'' of an isothermal was considered a triumph back in the 
1960's, when the
observed luminosity of many, if not all, elliptical galaxies could be well
fit by truncated isothermals, i.e., the so-called King models.$^{23}$
However, in the intervening years, improved high resolution photometry and
better background subtraction have yielded better data for which truncated
isothermals do not provide adequate fits. Instead, one is led typically to
model data using a de Vaucouleur$^{24}$ density profile or, more recently, a 
Nuker density law that allows for a central cusp.$^{25}$
\par
Even neglecting the fact that observed galaxies are not well fit
by truncated isothermals, the preceding doctrine of violent relaxation is open
to several criticisms. Perhaps the most obvious is simply: why should
one assume that ``most likely'' means (essentially) ``maximize the Boltzmann 
entropy''? The usual justification for this standard thermodynamic assumption, 
and for the combinatoric arguments exploited by Lynden-Bell, simply 
fails for a system interacting via a long range force like gravity. 
The standard argument relies crucially on the assumption of approximate 
extensivity, i.e., the assumption that, if one's system be viewed as a sum 
of individual pieces, quantities like the total energy are well approximated 
by a sum of the corresponding quantities for the individual pieces.$^{26}$
Indeed,
one could in principle play the same maximization game for any convex phase
space functional,$^{27}$ which leads to a natural connection with Casimir
arguments from plasma physics.$^{28}$
\par
More pragmatically, one infers from $N$-body simulations that a strongly
convulsing mean field potential is not necessary. One observes a comparably
efficient approach towards a meta-equilibrium on a time scale 
${\sim}{\;}t_{cr}$ both for ``violent'' evolution, where ${\Phi}$ exhibits
huge changes on very short time scales, and for ``nonviolent'' evolution, where
${\Phi}$ exhibits only relatively small changes.$^{29}$ Nonviolent relaxation
can be just as efficient as violent relaxation.
\par
Perhaps the most striking feature about violent relaxation is the assumption
that the evolution of ${\overline f}$ is completely unconstrained aside from
the imposition of some small number of bulk holonomic constraints and 
coarse-grained conservation of 
phase.\footnote{\small
As Lynden-Bell discussed in an Appendix to his paper, it is 
straightforward to generalise his derivation of eq.~(10), which only conserved
the mean field energy (6), to allow for conservation of the mean field angular 
momentum or any other quantity that can be written as a phase space integral.} 
This implies seemingly that, when 
viewed in terms of ${\overline f}$, the flow is chaotic, modulo only a 
restriction to a suitably defined phase space hypersurface on which bulk
quantities like the total energy are constant. 
\par
That this assumption is suspect was recognized already when 
violent relaxation was first proposed,$^{30}$ but at that time it was not
possible to implement direct tests computationally. However, during the past
fifteen years a number of different workers (cf. Refs. [31-33]) have exploited 
improved computational resources to test this assumption, and have been led
to the conclusion that it is {\it not} justified. Specifically, analyses of
$N$-body simulations have demonstrated that there is a precise sense in which
individual particles tend to ``remember'' certain aspects of their initial
conditions, at least statistically. Thus, e.g., for a variety of different
geometries, allowing for both collapse simulations and collisions between
galaxies, one discovers that particles that start with small binding energies
tend to end up with small binding energies, and visa versa. In the context
of the {\it CBE}, this would seem to indicate that phase space cells with
different initial energies do not become ``well shuffled.''
\par
This effect can be quantified both microscopically and mesoscopically. For
example, one can order the particles in a simulation in terms of their binding 
energies at different times and then compute quantities like the rank
correlation ${\cal R}(0,t)$ between the initial ordering and
the ordering at some later time $t$.$^{33}$ The net result of such 
investigations is that, even allowing for an extremely violent evolution, where
the virial ratio $2{\cal K}/{\cal V}$ changes by factors of two or more 
within a crossing time $t_{cr}$, ${\cal R}(0,t)>0.6$ at $t=20t_{cr}$, well
after the system appears to have settled down towards a meta-equilibrium.
\par
Alternatively, one can bin the particles in a simulation in terms of their 
initial binding energies, and then study the properties of the different bins 
as the system evolves into the future. Here the obvious question is whether, 
and if so to what extent, the statistical properties of particles in the 
initial low and high energy bins converge, as would be expected if the 
particle energies become completely randomized.$^{32,33}$ The conclusion of
such an analysis is that, at least over time scales ${\ll}{\;}t_{R}$, the mean 
energies of the different bins show absolutely no tendency to converge, 
although the energy dispersions of the different bins do grow.
\par
The seemingly unambiguous conclusion derived from these numerical 
investigations is that, contrary to what is usually assumed in violent
relaxation, the evolution towards a
meta-equilibrium can be comparatively nonviolent, and that, even for a violent
evolution, particle energies are not totally randomized. One cannot assume
a completely shuffled distribution of particle energies. There is thus a need
for an alternative theory of ``nonviolent relaxation''$^{32,34}$ which does
not assume a completely chaotic flow.
\vskip .12in
\centerline{\bf Alternatives to Violent Relaxation}
\vskip .08in
When considering a plasma which is displaced only slightly from some
meta-equilibrium $f_{0}$, it is customary to visualise evolution in terms
of linear Landau damping$^{35}$ which, for simple geometries, implies that
perturbations in the density and potential, ${\delta}{\rho}$ and 
${\delta}{\Phi}$,
damp exponentially. Physically, this damping is interpreted (cf. Ref. [36])
as reflecting a resonant coupling of (unperturbed) particles with physical 
velocity $v^{a}$ and a wave (the perturbation) with phase velocity $c^{a}$. 
Note that this picture necessarily involves a coarse-grained description
since it is formulated in terms of configuration space observables like
${\rho}$ and ${\Phi}$ rather than the full distribution function $f$. 
\par
It is difficult to interpret, let alone implement, Landau's original
calculations for systems that are strongly inhomogeneous. However, there is an
alternative interpretation which continues to make sense, namely that Landau 
damping involves the phase mixing of a superposition of normal modes.
In particular, the distinction which galactic dynamicists are wont to make
(cf. Ref. [2]) between phase mixing and Landau damping is simply not valid. 
Viewed appropriately, {\it linear Landau damping is a type of phase mixing.}
\par
A small perturbation ${\delta}f$ governed by the Vlasov equation appropriate
for a neutral electrostatic plasma will satisfy a
linearised evolution equation which can be written symbolically in the form
$${{\partial}{\delta}f\over {\partial}t}={\cal A}{\delta f}, \eqno(11) $$
where ${\cal A}$ is a linear integro-differential operator. Suppose, however,
that the initial perturbation ${\delta}f(0)$ is expanded in terms of the
normal modes of ${\cal A}$. If the spectrum of ${\cal A}$ is continuous, 
a smooth initial perturbation necessarily involves a superposition of normal
modes, but the evolution of this initial perturbation then implies that 
quantities like ${\delta}{\rho}$ and ${\delta}{\Phi}$ necessarily damp
exponentially.$^{37}$ In other words, Landau damping is guaranteed. This is 
intrinsically a phase mixing process, involving the nondissipative spreading
of an initial wavepacket. The perturbation manifests an initial coherence
that is lost as it evolves into the future. The full ${\delta}f$ itself does 
not tend towards zero in any smooth, pointwise sense, but coarse-grained 
observables like ${\delta}{\rho}$ and ${\delta}{\Phi}$ do.
\par
Alternatively, the existence of one or more discrete modes implies that there
are perturbations which do not phase mix or Landau damp. The classic example
thereof is van Kampen$^{38}$ modes, which correspond to waves executing 
undamped oscillations with a phase velocity $c^{a}$ where resonant coupling
to the unperturbed particles is impossible because 
$f_{0}(c^{a}){\;}{\equiv}{\;}0$.
\par
The important point here is that whether or not Landau damping/phase mixing
always occurs is determined completely by whether or not the spectrum of 
${\cal A}$ is purely continuous, and that the specific form of the modes is
irrelevant. For this reason, it is possible to prove an analogous result for
small perturbations of a gravitational {\it CBE} equilibrium.$^{39}$ 
Specifically,
if the modes of the linearised perturbation equation are all continuous,
any linearised perturbation will always exhibit Landau damping/phase mixing.
If, however, discrete modes exist, there are possible initial perturbations
which do not damp.
\par
Determining whether or not the normal modes for realistic {\it CBE} equilibria
are continuous is extremely difficult. In general it seems impossible to 
calculate the modes explicitly, and a formal analysis is also hard since 
${\cal A}$ is not an elliptic operator and evolves a singular integral kernel.
It is, however, worth recalling that, at least in a plasma, discrete modes
are typically associated with nontrivial boundary conditions, e.g., spatial
confinement or the existence of a maximum velocity. Thus, in particular, 
exponential 
damping is guaranteed if $f_{0}$ is analytic. However, one can argue that
self-gravitating {\it CBE} equilibria always have nontrivial boundary 
conditions. They are always confined to a compact spatial region by the
self-consistently determined potential; and there always exists a maximum
speed $|v_{max}(x^{a})|$ since characteristics for which $v$ is too large will
quickly escape to infinity.
\par
For this reason, one might reasonably conjecture$^{39}$ that many, but not
necessarily all, perturbations will Landau damp. Most initial 
${\delta}{\rho}$'s might be expected to damp, but undamped
perturbations could exist for modes that manifest coherence on a scale
comparable to the size $R_{sys}$ of the system, or, especially, given 
nontrivial phase space structures. In this regard, it seems significant that 
the aforementioned models of galaxies that exhibit undamped 
oscillations$^{21,22}$ all contain phase space ``holes,'' i.e., 
regions in the middle of the occupied phase space regions where $f_{0}\to 0$.
\par
The discussion hitherto has focused on linear Landau damping, i.e., the 
behaviour of a small initial perturbation which satisfies a linearised 
evolution equation. However, at least in principle this picture can also be 
extended to finite amplitude perturbations. In the context of plasma physics,
this means considering nonlinear Landau damping, which generalises the linear
theory by incorporating the possibility of mode-mode couplings that allow 
energy transfer between different modes. Because it is very difficult in a
realistic gravitational setting to determine the form of the normal modes even 
for a linearised perturbation, the development of a detailed theory of 
nonlinear gravitational Landau damping would be extremely difficult. However,
despite this problem it is still possible to visualise self-gravitating
systems in an identical fashion, even though the forms of the modes are not
known explicitly. 
\par
It would thus seem reasonable to {\it ask whether it is possible, 
and fruitful, to interpret the evolution of a generic initial $f(0)$ towards 
some meta-equilibrium as a manifestation of nonlinear Landau damping.} Such a
picture, based on a nonlinear evolution driven by mode-mode coupling, is by
construction a relatively coherent process, consistent with the statistical
regularities observed in $N$-body simulations and significantly less 
``incoherent'' than what is implicit in the standard picture of violent 
relaxation.
\par
Is there any evidence that this interpretation is actually viable? One
reassuring point is that, at least naively, one might expect the time scales to
work out correctly. Indeed, by analogy with Landau damping in a plasma, as
well as from simple dimensional analysis, one would expect that a perturbation
manifesting coherence on a scale ${\sim}{\;}R_{sys}$ should damp on a time
scale of order $t_{cr}$. Strictly speaking, this dimensional argument is only
justified for the linear theory, since mode-mode couplings could introduce
new time scales related to $t_{cr}$ by ratios of different length scales. 
However, at least for plasmas one can show that in certain cases allowing
for nonlinear mode-mode couplings does not lead to new time scales.$^{40}$
\par
The intuition that dynamical evolution can lead to undamped oscillating modes 
is of course corroborated by the aforementioned one-dimensional 
simulations of both gravitational and plasma systems with counter-streaming
initial conditions.$^{10}$ Here the late time state involves reasonably large
amplitude oscillations, corresponding to nearly periodic variations in the 
virial ratio $2{\cal K}/{\cal V}$ at the 10\% level, characterised by a period 
${\sim}{\;}t_{cr}$ and a coherence length ${\sim}{\;}R_{sys}$. Unlike the 
exact pulsating solutions,$^{21,22}$ which
incorporate true periodicity, the late time state arising in these 
one-dimensional simulations is not exactly periodic. Rather what one finds
is that the non-oscillating, nearly time-independent, component 
slowly changes as, via phase mixing (``phase wrapping''), it manifests a 
coarse-grained approach towards a truly time-independent $f_{0}$.
\par
Another point, not conclusive but arguably suggestive, arises from a 
consideration of moment codes. These codes approximate the Vlasov equation
of plasma physics or the {\it CBE} of gravity by a discrete collection of 
coupled moment equations which have been reduced to a finite set through the 
introduction of a symplectic trunction (for example, Channell's$^{41}$
{\it BEDLAM} code for an electrostatic plasma and the corresponding 
{\it GADFLI} for a self-gravitating system each involve a collection of
$203$ coupled moments). When used by accelerator dynamicists to study the 
stability of beam flows, these efficient moment codes have proven 
spectacularly successful in the sense that they agree with theory and 
$N$-body simulations (and experiments!), and one might expect analogously
that their gravitational analogues could prove powerful tools in diagnosing
stability or lack thereof for solutions to the {\it CBE}. 
\par
However, when applied to a system that is far from equilibrium {\it GADFLI} 
(GAlactic Dynamics From Lie-Poisson Integration) yields results that differ 
markedly from what is observed in $N$-body simulations.$^{42}$ Specifically, 
{\it GADFLI} does not yield the comparatively efficient approach towards 
equilibrium that is observed in simulations. For example, whereas $N$-body
simulations of a collapsing spherical configuration typically exhibit 
oscillations which damp on a time scale ${\sim}{\;}t_{cr}$, the moment code
yields oscillations that only damp appreciably on a significantly longer
time scale. The obvious inference is that {\it GADFLI} is missing some 
important physical ingredient that is present in $N$-body codes. One possible 
interpretation$^{43}$ would be that standard $N$-body
codes are contaminated by non-symplectic perturbations that act as a spurious
source of dissipation, and that a more satisfactory $N$-body code would
exhibit a less efficient approach towards equilibrium. 
($N$-body codes that use a time-centered leap frog scheme are in fact 
symplectic, but this scheme is not usually used for large $N$ simulations 
which tend, generically, to exploit variable time steps.) However, a more
conservative interpretation is that the $N$-body codes are at least 
approximately correct, and that {\it GADFLI} misses the effects of phase mixing
which rely on the existence of a large number -- formally an infinite number
-- of modes.
\par
If one accepts tentatively the hypothesis that the flow associated with a 
self-gravitating system far from equilibrium should be interpreted in terms of 
nonlinear Landau damping, one is led to a single, unified picture of an 
evolution described by the {\it CBE} in which the single dominant mechanism
is phase mixing/Landau damping. The idea is that, for a generic initial
$f(0)$, the approach towards (albeit not to) an equilibrium will involve
nonlinear oscillations about some time-independent $f_{0}$, oscillations
which may -- or may not -- eventually phase mix away. When the system is far
from equilibrium, nonlinearities will be important and mode-mode coupling 
will play an important role. Eventually, however, as the system approaches
equilibrium more closely, nonlinearities will become less important and 
ordinary linear Landau damping/phase mixing will provide a reasonable
description of what is actually happening.
\par
Given this hypothesis, it is also clear what problem one would like ideally
to solve: Determine how to break a generic $f(0)$ into two pieces, an
equilibrium $f_{0}$ and a correction ${\delta}f(0)$ not necessarily small,
such that, at least in the absence of any undamped modes, the perturbation 
${\delta}{\rho}$ and/or ${\delta}{\Phi}$ associated with the evolved 
${\delta}f$ tends to zero at late times.
\par
Another, slightly different, way in which to formulate this same picture,
which may provide some insights into how to identify the equilibrium
$f_{0}$ lurking in the initial $f(0)$, is the following: A Hamiltonian
evolution governed by the {\it CBE} corresponds to a flow in an 
infinite-dimensional phase space which is constrained by conservation of
phase (i.e., fixed values of all the Casimirs $C$)
and the fact that the value of the mean field energy ${\cal H}$
does not change in time. Imposing these constraints explicitly serves to 
pick out a (still infinite-dimensional) hypersurface in the phase space to
which the flow is restricted. By analogy with finite-dimensional Hamiltonian
systems (cf. Refs. [44,45]), one might expect that, when evolved into the 
future, an initial $f(0)$ will exhibit a coarse-grained approach towards 
some invariant measure on this hypersurface, e.g., towards a suitably defined 
microcanonical population of the accessible phase space region. If, when 
viewed in the infinite-dimensional
phase space, the flow is chaotic, so that nearby distribution functions
$f$ diverge exponentially, one might expect an approach towards an invariant
measure which is exponential in time. If, alternatively, the flow is regular,
one might instead expect a power law approach. However, in either case it
would seem reasonable to expect an approach towards a ``phase-mixed'' 
microcanonical distribution proceeding on the natural time scale $t_{cr}$.
In this context, the crucial questions would seem to be the following: 
for the fixed values of $C$ associated with $f(0)$, what equilibrium solutions 
$f_{0}$ exist, and to what extent can the invariant distribution associated 
with the evolved $f(0)$ be interpreted as involving (nonlinear)
oscillations about some lower energy $f_{0}$?
\vskip .1in
\centerline{\bf Weak Non-Hamiltonian Corrections to Collisionless
Evolution}
\vskip .1in
This entire picture relies crucially on the fact that the {\it CBE} is a 
Hamiltonian
system. That the {\it CBE} is Hamiltonian is in turn a manifestation of a very
general result, namely that, when suitably interpreted, the mean field 
description of {\it any} Hamiltonian system is itself Hamiltonian.$^{46}$
If, however, any corrections to mean field theory are introduced, one obtains
generically a description of the one-particle distribution function $f$ which
is no longer Hamiltonian. Physically, this arises because one is treating the
full $N$-particle distribution function ${\mu}$  as being a sum of two 
``pieces,'' namely (1) a product of identical one-particle distribution 
functions $f(i)$ ($i=1,...,N$) and (2) a correction reflecting 
particle-particle correlations
which are ignored in the mean field description. The point then is that 
integrating over the correlations to obtain a closed equation for $f$ leads 
to a nonlocal, non-Hamiltonian evolution equation.$^{47}$ 
\par
At the level of particle trajectories, this intuition is manifested in the
conventional description of binary encounters as resulting in diffusion and
dynamical friction.$^{2,6}$ Thus, e.g., for motion in a fixed potential, one
can formulate evolution equations of the form
$${dx^{a}\over dt}=v^{a} \qquad
{\rm and} \qquad
{dv_{a}\over dt}=-{{\partial}{\Phi}\over {\partial}x^{a}} 
-{\eta}(x^{a},v_{a})v_{a}+{\cal F}_{a}, \eqno(12) $$
where ${\eta}$ and ${\cal F}_{a}$ are related by a fluctuation-dissipation
theorem, in terms of a characteristic ``temperature'' or mean squared velocity
${\Theta}$. If, for example, one treats the fluctuating forces as Gaussian 
white noise with zero mean, ${\eta}$ must be constant and (cf. Ref. [48])
$$ {\langle}{\cal F}_{a}(t_{1}){\cal F}_{b}(t_{2}){\rangle}=
2{\Theta}{\eta}{\delta}_{ab}{\delta}_{D}(t_{1}-t_{2}). \eqno(13) $$
\par
If the friction and noise are weak, they will only cause appreciable
changes in the particle energy $E$ (and any other quantities conserved in
the Hamiltonian description) on a time scale ${\gg}{\;}t_{cr}$. However, 
this does not mean that they cannot have appreciable effects on the flow over
much shorter time scales. Indeed, they can be important in at least two ways:
\begin{enumerate}
\item \vskip -.10in
If the fixed potential is complicated, involving a coexistence of both
regular and chaotic orbits, a strictly Hamiltonian evolution can result in
chaotic orbits being trapped in certain phase space regions for a comparatively
long time, even though, in the $t\to\infty$ limit, they will diffuse through
cantori$^{49}$ or along an Arnold web$^{50}$ to probe a much larger phase
space region.$^{51,52}$ However, even very weak friction and noise -- so weak
that, over short time scales, they have negligible effects on the form of
the invariant measure -- tend generically to dramatically accelerate this 
diffusion, thus facilitating a much more rapid approach towards the invariant 
measure.$^{52-54}$ Thus, e.g., in some cases friction and noise corresponding
to a natural time scale as long as $10^{6}-10^{9}t_{cr}$ can have significant 
statistical effects on the evolution of ensembles of orbits on a time scale
as short as $100t_{cr}$. At least for orbits in a fixed potential, small 
non-Hamiltonian corrections can greatly facilitate the coarse-grained approach 
towards a statistical equilibrium.
\item \vskip -.10in
Liouville's Theorem implies that trajectories in a Hamiltonian system cannot
self-intersect. Thus, no matter how complicated the potential or the phase
space, a Hamiltonian trajectory must evolve in such a fashion as to avoid
crossing itself. However, this can lead to complicated microscopic
structures, such as, e.g., the ``homoclinic tangle'' that can arise near
separatrices.$^{55}$ The important point then is that even very weak 
non-Hamiltonian
perturbations can break the constraints associated with Liouville's Theorem,
thus allowing the tangle to smooth itself out. In other words, small 
perturbations that break the rigid constraints associated with Liouville's
Theorem can serve to fuzz out small scale structures that arise in a purely
Hamiltonian description.
\end{enumerate}
\par
These two alternatives have been formulated completely in the context of motion
in a fixed potential; and, at the present time, little if anything is really
known about the effects of small non-Hamiltonian perturbations in the context
of a self-consistent description as provided by the {\it CBE}. However, by 
analogy with the simpler problem it would seem reasonable to conjecture that
even very weak perturbations that break the constraints associated with 
conservation of phase could facilitate the approach towards a meta-equilibrium,
and that they could help blur the complex structures which can arise
microscopically as the system phase mixes towards this meta-equilibrium.
\vskip .2in
\centerline{\bf HOW DOES THE COLLISIONLESS BOLTZMANN EQUATION}
\vskip .05in
\centerline{\bf RELATE TO THE FULL N-BODY PROBLEM?}
\vskip .15in
The gravitational $N$-body problem for a collection of particles of comparable 
mass
is chaotic in the sense that a small initial perturbation will typically grow
exponentially on a time scale ${\sim}{\;}t_{cr}$. This phenomenon, long 
known,$^{56}$ has been studied systematically over the past several 
years,$^{57-61}$ and the observed behaviour has proven to be extremely 
robust. Specifically, one finds that the observed exponential growth is largely
independent of the detailed choice of initial conditions and initial 
perturbation. Seemingly independent of the details, small initial perturbations
grow exponentially until they become ``macroscopic.'' Thus, in particular,
this chaos persists even for a system that is in or near a statistical
meta-equilibrium. Moreover, one finds that, when scaled in terms of an 
$N$-dependent crossing time $t_{cr}$, the time scale ${\tau}$ associated with 
this instability admits at most a weak dependence on $N$, at least for 
$N{\;}{\gg}{\;}2$. In particular, the chaos does not appear to turn off for 
very large $N$. There is no sense in which $({\tau}/t_{cr})\to\infty$ for 
$N\to\infty$. If anything, the ratio ${\tau}/t_{cr}$ very slowly 
{\it decreases} with increasing $N$.$^{60}$
\par
This leads, however, to an important question of principle. The $N$-body
problem appears to be chaotic on a time scale ${\sim}{\;}t_{cr}$, but the
flow associated with the {\it CBE} is often integrable or near-integrable in
the sense that many/all of the characteristics are regular, i.e., nonchaotic. 
So what do the (often near-integrable) {\it CBE} characteristics have to
do with the true (chaotic) $N$-body problem?
\par
There is no {\it a priori} contradiction between a chaotic $N$-body problem
and a {\it CBE} evolution that is integrable or near-integrable, since the
{\it CBE} is predicated on the existence of a smooth distribution function
$f$ and a smooth potential ${\Phi}$. However, there is at least one awkward
point: Although galactic astronomers teach that {\it CBE} characteristics
simply track the motion of phase space fluid elements, and that they should
not be interpreted as representing the orbits of representative particles,
there {\it is} oftentimes the implicit idea that, for $N\to\infty$, true
$N$-body orbits will converge towards orbits in the smooth, self-consistent
potential, i.e., towards {\it CBE} characteristics.
\par
For a singular initial distribution function
$$f_{N}(x^{a},v_{a},0)=\sum_{i=1}^{N}\,m\,{\delta}_{D}[{\bf x}-{\bf x}_{i}(0)] 
{\delta}_{D}[{\bf v}-{\bf v}_{i}(0)], \eqno(14) $$
the {\it CBE} is of course equivalent to the full $N$-body problem. If, 
however, one considers a sequence of initial distribution functions with
increasing number $N$, decreasing particle mass $m$, and constant $M=Nm$, 
there is no obvious
sense in which, for $N\to\infty$, the initial $f_{N}(0)$  converges towards
a smooth distribution function $f_{\infty}(0)$. Rather, what one has is lower
amplitude (but still infinite) spikes on progressively shorter scales. It is
therefore hard to envision a uniform convergence towards a smooth potential or
smooth {\it CBE} characteristics. 
\par
In this general context, it is also easy to understand qualitatively why
chaos might be expected to persist even for very large $N$. Numerical 
simulations suggest that much, if not all, of the chaos observed in the 
$N$-body problem is associated with ``random'' interactions between nearest
neighbours, perhaps amplified through a coupling to the bulk mean field.
If, however, this is true, the time scale associated with the growth of
an initial perturbation can be estimated by considering the tidal effects
associated with the perturbation of a pair of particles separated by a 
distance comparable to a typical interparticle spacing. The tidal acceleration
associated with a given pair will of course scale as
$${\delta}{\ddot {\bf x}}=({\delta}{\bf x}{\cdot}{\nabla}){\bf a}{\;}{\sim}{\;}
{Gm\over r^{3}}{\,}{\delta}{\bf x}, \eqno(15) $$
where $r$ is their separation and $m$ a typical particle mass. Assuming,
however, that $r{\;}{\sim}{\;}n^{-1/3}$, with $n$ a typical number density,
and that, at least roughly, $GNm/R_{sys}{\;}{\sim}{\;}{\overline v}^{2}$
with ${\overline v}$ a typical particle speed, it follows from 
dimensional analysis that the natural time scale 
${\tau}{\;}{\sim}{\;}R_{sys}/{\overline v}{\;}{\sim}{\;}t_{cr}$. As $N$ 
increases,
the mass of the nearest neighbour decreases, but the distance to that neighbour
also decreases in a fashion which cancels the effect of decreasing mass.
\par
This leads to another awkward point. $N$-body simulators often ``soften''
their computations to reduce discreteness effects which, presumably, are more
important in the small $N$ systems which can actually be solved than in the
real, large $N$ systems which one would like to understand. Thus, e.g., it is
customary in direct summation codes to replace the true $1/r$ kernel of 
Newtonian gravity by a new $1/\sqrt{r^{2}+{\epsilon}^{2}}$ kernel which
effectively turns off encounters on scales shorter than the softening length
${\epsilon}$. However, if ${\epsilon}$ is chosen to be too large, i.e.,
not much smaller than the typical interparticle spacing, such a softening
effectively kills the chaos which one expects to be present in the honest
$N$-body problem even for very large $N$.
\par
The correct answer to the question raised above -- how to reconcile a chaotic
$N$-body problem with a non-chaotic {\it CBE} flow -- is not completely clear. 
What does, however, seem apparent from the preceeding is that, even for very 
large $N$, true $N$-body trajectories could differ significantly from 
{\it CBE} characteristics.
\par
One interpretation of the {\it CBE}, which may enable one to finesse this
question, is an ``ensemble average'' interpretation (cf. Ref. [62,63]), in
which any given realisation of the $N$-body problem would be interpreted as
involving the evolution of a collection of $N$ initial conditions that
randomly sample the smooth distribution function $f(0)$. In this context,
there is no reason to expect that the orbits in an $N$-body realisation will
closely track characteristics of the {\it CBE}, but one might at least hope 
that, in some statistical sense, there is a close correspondence between 
$N$-body orbits and {\it CBE} characteristics. 
\par
To help focus on the basic issue, it is useful to consider an interesting
thought experiment which, given recent advances in computer hardware, could
perhaps actually be performed numerically. Start by specifying a smooth
distribution function $f(0)$ and computing the characteristic associated with
some phase point $({\bf x}_{0},{\bf v}_{0})$. Next sample $f(0)$ to generate
many different sets of initial conditions for the $N$-body problem, all with
one particle at $({\bf x}_{0},{\bf v}_{0})$, and evolve these different sets
of initial conditions into the future. The obvious question is then: How do
the $N$-body orbits generated from $({\bf x}_{0},{\bf v}_{0})$ compare with
the {\it CBE} characteristic, both individually and in terms of statistical
averages?
\par
Given the fact that the $N$-body problem is chaotic on a time scale 
${\sim}{\;}t_{cr}$, it would seem reasonable to conjecture that the orbits
generated in two different $N$-body realisations will diverge exponentially
on a time scale ${\sim}{\;}t_{cr}$; and, similarly, one might expect that
any given $N$-body orbit will diverge from the {\it CBE} characteristic on
a comparable time scale. However, one might nevertheless expect that, for
sufficiently large $N$, the ensemble average of the different $N$-body orbits 
generated from the same $({\bf x}_{0},{\bf v}_{0})$ will closely track the 
{\it CBE} characteristic for some finite time. In particular, one might 
conjecture that the {\it rms} configuration space deviation between the 
$N$-body orbits 
and the {\it CBE} characteristic will scale as
$${\delta}r_{rms}(t){\;}{\sim}{\;}F(N){\exp}(t/{\tau}) , \eqno(16) $$
where ${\tau}{\;}{\sim}{\;}t_{cr}$, roughly independent of the total particle
number $N$, and where the prefactor $F(N)\to 0$ for $N\to\infty$. Thus, e.g., 
the assumption of random fluctuations associated with a random choice of 
initial conditions might suggest $F(N){\;}{\propto}{\;}1/\sqrt{N}$.
\par
Whether these expectations are correct is not clear. What does, however, seem
evident is that galactic astronomers lack a clear understanding of the sense,
if any, in which $N$-body orbits coincide approximately with {\it CBE}
characteristics. This, however, is arguably an important lacuna. Much of the 
theorist's intuition derives from the study of orbits in smooth potentials
presumably associated with {\it CBE} equilibria; and much of the standard
interpretation of observational data is predicated on the assumption that
stars follow simple nonchaotic orbits in a smooth gravitational potential.
\vskip .2in
\centerline{\bf ACKNOWLEDGMENTS}
\vskip .15in
The ideas described here have derived in part from discussions -- and 
arguments -- with a number of different individuals, including George
Contopoulos, Paul Channell, Salman Habib, Ted Kirkpatrick, Phil Morrison, and
Daniel Pfenniger. I am grateful to Tom O'Neil, Dave Merritt, and Richard 
Lovelace for critical feedback on the talk from which this paper derives. I
am especially grateful to Donald Lynden-Bell for his detailed criticisms of
the penultimate version of the manuscript. 
The final draft was completed while 
I was a visitor at the Aspen Center for Physics.
\vfill\eject
\centerline{\bf REFERENCES}
\vskip .15in
\par\noindent
1. MIHALAS, D. \& BINNEY, J. 1981, Galactic Astronomy. Freeman, San Francisco.
\par\noindent
2. BINNEY, J. \& TREMAINE, S. 1987. Galactic Dynamics. Princeton University
Press, \par Princeton.
\par\noindent
3. SCHWEIZER, F. 1986. Science {\bf 231}: 227.
\par\noindent
4. MAOZ, E. 1991, Astrophys. J. {\bf 375}: 687.
\par\noindent
5. LYNDEN-BELL, D. 1967, Mon. Not. R. Astr. Soc. {\bf 136}: 101.
\par\noindent
6. CHANDRASEKHAR, S. 1943, Principles of Stellar Structure. University of
Chicago \par Press, Chicago.
\par\noindent
7. KANDRUP, H. E. 1981, Astrophys. J. {\bf 244}: 316.
\par\noindent
8. HERTEL, P. \& THIRRING, W. 1971. Commun. Math. Phys. {\bf 24}: 22.
\par\noindent
9. RYBICKI, G. 1971. Astrophys. Space Sci. {\bf 14}: 56.
\par\noindent
10. MINEAU, P., FEIX, M. R. \& ROUET, J. L. 1990. Astron. Astrophys. {\bf 228}:
344.
\par\noindent
11. LUWEL, M. \& SEVERNE, G. 1985. Astron. Astrophys. {\bf 152}: 305.
\par\noindent
12. YAWN, K. R. \& MILLER, B. N. 1997. Phys. Rev. E, in press.
\par\noindent
13. MORRISON, P. J.  1980, Phys. Lett. A {\bf 80}: 383.
\par\noindent
14. KANDRUP, H. E. 1990. Astrophys. J. {\bf 351}: 104.
\par\noindent
15. MORRISON, P. J. \& ELIEZER, S. 1986. Phys. Rev. A {\bf 33}: 4205.
\par\noindent
16. BARTHOLOMEW, P. 1971. Mon. Not. R. Astr. Soc. {\bf 151}: 333.
\par\noindent
17. PFAFFELMOSER, K. 1992. J. Diff. Eqns. {\bf 95}: 281.
\par\noindent
18. SCHAEFFER, J. 1991. Commun. Part. Diff. Eqns. {\bf 16}: 1313.
\par\noindent
19. KORMENDY, J. \& RICHSTONE, D. 1995. Ann. Rev. Astron. Astrophys. {\bf 33}:
581.
\par\noindent
20. BATT, J. 1987. Transport Theory and Statistical Physics {\bf 16}: 763.
\par\noindent
21. LOUIS, P. D. \& GERHART, O. E. 1988. Mon. Not. R. Astr. Soc. {\bf 233}:
337.
\par\noindent
22. SRIDHAR, S. 1989. Mon. Not. R. Astr. Soc. {\bf 201}: 939.
\par\noindent
23. KING, I. 1962, Astron. J. {\bf 67}: 471.
\par\noindent
24. DE VAUCOULEUR, G. 1948, Ann. d'Astrophysique {\bf 11}: 247.
\par\noindent
25. LAUER, T. R., AJHAR, E. A., BYUN, Y.-L., DRESSLER, A., FABER, S. M.,
GRILL-
\par
MAIR, C., KORMENDY, J., RICHSTONE, D., \& TREMAINE, S. 1995. Astron. J. 
\par
{\bf 110}: 2622.
\par\noindent
26. LANDAU, L. D. \& LIFSHITZ, E. M. 1958. Statistical Physics. Pergamon,
London.
\par\noindent
27. IPSER, J. R. \& HORWITZ, G. 1979. Astrophys. J. {\bf 232}: 863.
\par\noindent
28. HOLM, D. D., MARSDEN, G. E., RATIU, T. \& WEINSTEIN, A. 1985. Phys. 
Repts. \par {\bf 123}: 1.
\par\noindent
29. KANDRUP, H. E., MAHON, M. E., \& SMITH, H. 1993. Astron. Astrophys.
{\bf 271}: \par 440.
\par\noindent
30. CONTOPOULOS, G. 1994. private communication.
\par\noindent
31. VAN ALBADA, T. S. 1982, Mon. Not  R. Astr. Soc. {\bf 201}: 939.
\par\noindent
32. QUINN, P. J. \& ZUREK, W. H. 1988. Astrophys. J. {\bf 331}: 1.
\par\noindent
33. KANDRUP, H. E., MAHON, M. E., \& SMITH, H. 1993. Astron. Astrophys. 
{\bf 271}; 440.
\par\noindent
34. KANDRUP, H. E. 1993. {\it In} Advances in Gravitation and Cosmology.
B. R. Iyer, A. R. 
\par
Prasanna, R. K. Varma, and C. V. Vishveshwara, Ed. Wiley
Eastern, New Delhi.
\par\noindent
35. LANDAU, L. D. 1946. Soviet Phys. - JETP {\bf 10}: 25.
\par\noindent
36. STIX, T. H. 1962, The Theory of Plasma Waves. McGraw-Hill, New York.
\par\noindent
37. CASE, K. M. 1959. Ann. Phys. (NY) {\bf 7}: 349.
\par\noindent
38. VAN KAMPEN, N. G. 1955, Physica {\bf 21}: 949.
\par\noindent
39. HABIB, S., KANDRUP, H. E., \& YIP, P. F. 1986, Astrophys. J. {\bf 309}:
176.
\par\noindent
40. MONTGOMERY, D. 1963, Phys. Fluids A {\bf 6}: 1109.
\par\noindent
41. CHANNELL, P. J. 1995. Ann. N. Y. Acad. Sci. {\bf 751}: 152.
\par\noindent
42. KANDRUP, H. E. 1994. Is there Chaos in BEDLAM? University of Florida
preprint.
\par\noindent
43. CHANNELL, P. J. 1994. private communication.
\par\noindent
44. KANDRUP, H. E. \& MAHON, M. E. 1994. Phys. Rev. E {\bf 49}: 3735.
\par\noindent
45. MAHON, M. E., ABERNATHY, R. A., BRADLEY, B. O., \& KANDRUP, H. E. 1995.
\par Mon. Not. R. Astr. Soc. {\bf 275}: 443.
\par\noindent
46. KANDRUP, H. E. 1994. Phys. Rev. D {\bf 50}: 2425.
\par\noindent
47. KUBO, R., TODA, M., \& HASHITSUME, N. 1978, Statistical Physics II.
Springer-
\par
Verlag, New York. 
\par\noindent
48. CHANDRASEKHAR, S. 1943. Rev. Mod. Phys. {\bf 15}: 1.
\par\noindent
49. MATHER, J. N. 1982, Topology {\bf 21}: 45.
\par\noindent
50. ARNOLD, V. I. 1964, Russian Math. Surveys {\bf 18}: 85.
\par\noindent
51. CONTOPOULOS, G. 1971. Astron. J. {\bf 76}: 147.
\par\noindent
52. LIEBERMAN, M. A. \& LICHTENBERG, A. J. 1972. Phys. Rev. A {\bf 5}: 1852.
\par\noindent
53. HABIB, S., KANDRUP, H. E., \& MAHON, M. E. 1996. Phys. Rev. E {\bf 53}: 
5473.
\par\noindent
54. HABIB, S., KANDRUP, H. E., \& MAHON, M. E. 1997. Astrophys. J., in press.
\par\noindent
55. LICHTENBERG, A. J. \& LIEBERMAN, M. A. 1992, Regular and Chaotic Dynamics.
\par
Springer-Verlag, New York.
\par\noindent
56. MILLER, R. H. 1964, Astrophys. J. {\bf 140}: 250.
\par\noindent
57. KANDRUP, H. E. \& SMITH, H. 1991. Astrophys. J. {\bf 374}: 255.
\par\noindent
58. KANDRUP, H. E. \& SMITH, H. 1992. Astrophys. J. {\bf 386}: 635.
\par\noindent
59. KANDRUP, H. E., SMITH, H., \& WILLMES, D. E. 1992. Astrophys. J. {\bf 399}:
 627.
\par\noindent
60. GOODMAN, J., HEGGIE, D., \& HUT, P. 1994. Astrophys. J. {\bf 415}: 715.
\par\noindent
61. KANDRUP, H. E., MAHON, M. E., \& SMITH, H. 1992. Astrophys. J. {\bf 428}: 
458.
\par\noindent
62. OGORODNIKOV, K. F. 1965. Dynamics of Stellar Systems. MacMillan, New York.
\par\noindent
63. KLIMONTOVICH, YU. L. 1983, The Kinetic Theory of Electromagnetic Processes.
\par Springer-Verlag, New York.
\par\noindent
64. LYNDEN-BELL, D. \& WOOD, R. 1968, Mon. Not. R. Astr. Soc. {\bf 138}: 495.
\eject\end{document}